\documentclass[prb,aps,amsmath,preprint]{revtex4}
\usepackage{graphicx}
\usepackage{dcolumn}
\usepackage{bm}
\usepackage{subfigure}
\usepackage{scrextend}
\usepackage{hyperref}
\usepackage{color}

\begin{document}
\title{Structural, vibrational, and electronic properties of TlReO$_4$ studied using a first principles approach}
\author{S. Mondal}
\affiliation{Advanced Centre of Research in High Energy Materials (ACRHEM),\\
University of Hyderabad, Prof. C. R. Rao Road, Gachibowli, Hyderabad - 500 046, Telangana, India.}
\author{G. Vaitheeswaran$^*$}
\affiliation{School of Physics, University of Hyderabad, Prof. C. R. Rao Road, Gachibowli, Hyderabad-500046, Telangana, India.}
\email{vaithee@uohyd.ac.in (Corresponding Author)}
\author{Daniel Errandonea$^{\dagger}$}
\affiliation{Departamento de Fisica Aplicada-ICMUV-MALTA Consolider Team, Universidad de Val\`{e}ncia, C/Dr. Moliner 50, 46100 Burjassot, Valencia, Spain.}
\email{daniel.errandonea@uv.es (Corresponding Author)}
\date{\today}

\begin{abstract}
The present work provides an extensive density functional theory study on the ambient phase of TlReO$_4$. In this connection, a pseudo-potential approach has been considered for understanding the structural and vibrational properties of this material whereas the investigation of electronic band structure and associated optical properties of this material has been carried out through the implementation of full-potential linearized augmented plane wave (FP-LAPW) method. The calculated bulk modulus value 29.77 GPa using LDA functional is found
to be close to experimental value 26 GPa. The order of elastic constant along the principal axis: C$_{22}$ $>$ C$_{33}$ $>$ C$_{11}$, clearly
indicates that the Re$-$O3 bonds play a crucial role in the structural and mechanical properties of this material. An analysis of the Born effective charge (BEC) along with $\Gamma$-point phonon frequencies through density functional perturbation approach (DFPT) have also shown the importance of the Re$-$O3 bond. The anisotropic nature of BEC is mainly found to be contributed due to the O3 atoms. An asymmetric stretching of the Re$-$O3 bond is found to be mainly responsible for high intense IR peak in the high-frequency range whereas the second most intense peak is due to the symmetric stretching of these bonds along with the asymmetric stretching of the Re$-$O2 bonds. In order to get the exact electronic band structure, spin orbit coupling (SOC) in addition to Tran-Blaha Modified Becke-Johnson (TB-mBJ) potential have been considered. Inclusion of SOC clearly shows a decreased band gap of 4.71 eV from TB-mBJ implemented band gap of 4.82 eV which is mainly attributed to a prominent splitting of Re \emph{d}-states of about 0.08 eV. Study of density of states reveals that the conduction band bottom (CBB) is mainly made up of Re \emph{d}-states but the main contribution to valence band top (VBT) is due to oxygen \emph{p}-states. The bonding nature of this material has also been addressed using density of state (DOS) which is further verified through the electron charge density plot. Though the optical properties of this material are found to be anisotropic but optical isotropy can be seen in lower energy value specially in XX and YY directions which could be due to O2 and O3 type oxygen atoms.
\end{abstract}
\maketitle

\section{Introduction}
ABO$_4$ type oxides with different crystal structures are not only important in Earth Science,\cite{1} and Material Science \cite{2,3,4} but also
the study of these materials is worth mentioning from the Physics point of view\cite{5}. A wide variety of these oxides are mainly found to
crystallize in zircon (silicates, phosphates, arsenates, vanadates and chromates), quartz (phosphates, arsendates), scheelite (germanates, molybdates, tungstates and periodates)\cite{6}, wolframite (molybdates, tungstates and tantalates), M-fergusonite (distorted scheelite structure) and pseudoscheelite-type\cite{7} crystal structures at ambient conditions\cite{8}. Among these oxides, ABO$_4$ type CdWO$_4$ and PbWO$_4$ are known as scintillator crystals, ZrGeO$_4$ and HfGeO$_4$ are known for their use as phosphors, BaWO$_4$ and GdTaO$_4$ are considered as laser host materials and CaMoO$_4$ and SrWO$_4$ are known for their use in batteries\cite{9}. Report on germanate materials (ZrGeO$_4$ and HfGeO$_4$) shows that the doping of Ti as impurity will make these materials better phosphors than pure compounds which in turn enhance the technological applications\cite{10}. In case of scintillator crystals, orthovanadates have shown a better performance than periodates at ambient as well as at high pressure\cite{11,12,13} and a recent finding of optical isotropy in anisotropic perchlorate structures clearly suggests its use as inorganic scintillator\cite{14}. Apart from different useful applications, most of these ABO$_4$ type materials are also known to exist in different phases and exhibit pressure-induced phase transition. In order to understand this phenomenon of ABO$_4$ type materials, ionic size and the valence of cations A and B have been the main focus in theoretical study of such materials\cite{15,16}. The stability of these structures with respect to different cations and the study of pressure-induced phase transition upon varying pressure is the idea of this kind of study. To have a proper understanding of the stability and the pressure induced phase transition of these structures, experimental studies using Raman spectroscopy have been performed. Among these studies, the investigation of the stability of ABO$_4$ scheelite-type tungstates and molybdates upon compression is worth mentioning\cite{16,17}. AReO$_4$ perrhenates, with A as alkali metal, ammonium, or Tl are also reported to crystallize in the scheelite or pseudoscheelite crystal structure and undergo phase transitions\cite{18} upon applied pressure or temperature.\\
In connection with it, TlReO$_4$ has got a great attention for the possible structural as well as electronic phase transition at high pressure. X-ray diffraction\cite{19,20}, x-ray near-edge absorption spectroscopy\cite{21}, and Raman spectroscopy\cite{22} have been reported on this material. These studies suggest that TlReO$_4$ crystallizes in the pseudoscheelite structure at ambient conditions\cite{18} and upon heating to 400 K and cooling to 200 K two phase transitions were observed\cite{20,24}. Pressure-induced phase transitions have also been noticed at 0.5, 1.9 and 9.7 GPa\cite{22,25}. Among them, the phase transition at 9.7 GPa triggers a color change of the material\cite{22}. An electronic transition mechanism has been proposed to understand this phenomenon.\\
Though these experiments give a good amount of information about the phase transition of TlReO$_4$, little information is known of the main physical properties of this compound. In particular, there is a lack of theoretical studies on TlReO$_4$, which are fundamental for the interpretation of experiments. It has been recently shown that a systematic theoretical calculation based on DFT would provide deep insight into the physical properties of ABO$_4$-type oxides\cite{23}, but this has not been carried out yet for TlReO$_4$. As a first step to achieve this goal, a proper understanding of different properties such as structural, vibration and optical properties of this material at ambient phase has been addressed in this article. The good agreement obtained on the crystal structure and phonon frequencies make us confident on the predictions made for other properties like band-gap and refractive index.\\
This article is organized as follows: crystal structure along with computational details of this calculation has been discussed in the next section. After that, the results and discussion on structural, mechanical, electronic, optical, and dynamical properties have been made. Finally, the conclusion of this study will be presented.

\section{Crystal Structure and Computational details}
\begin{figure}
 \subfigure[]{\includegraphics[width=3.3in,clip]{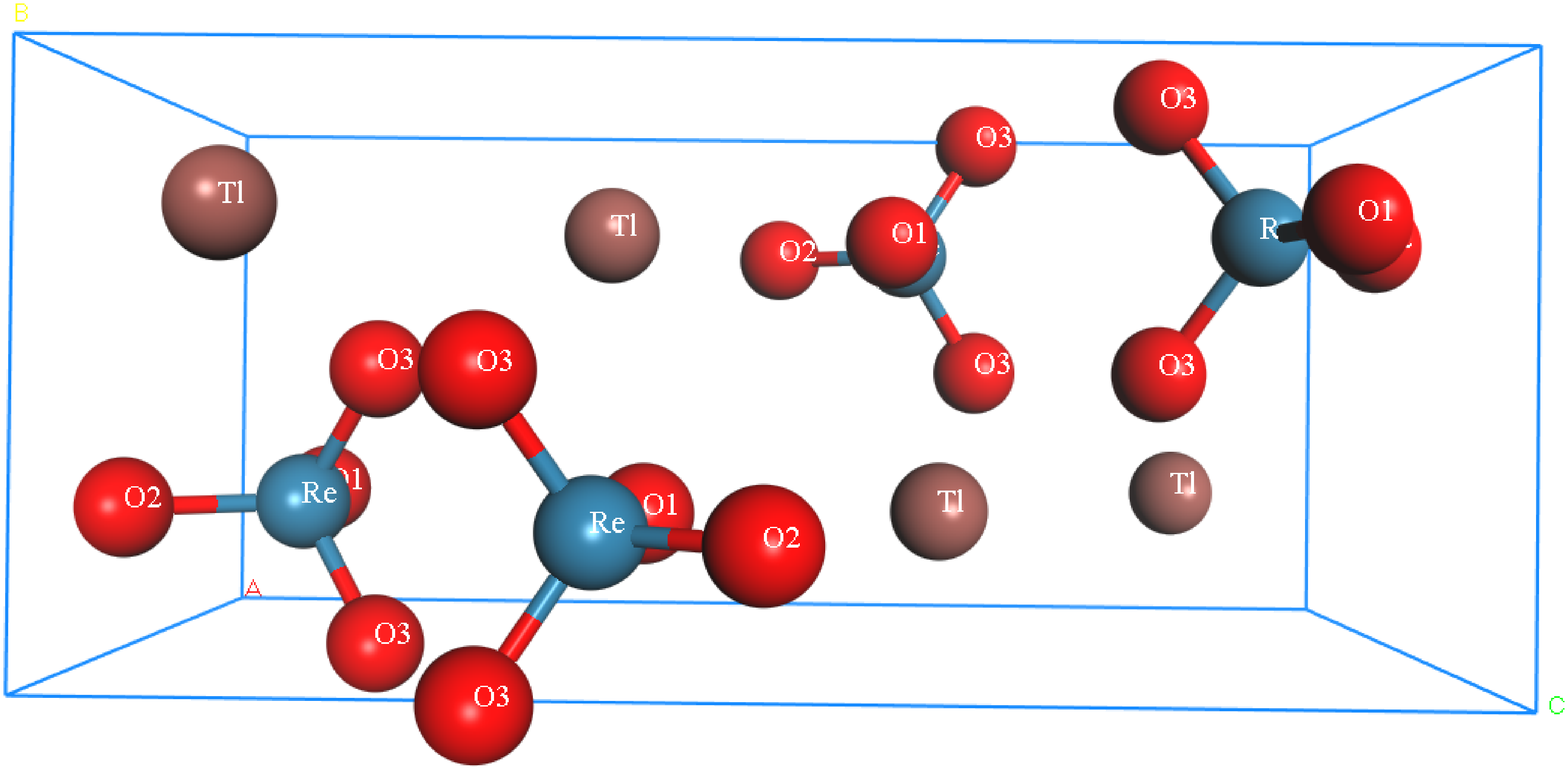}}
 \subfigure[]{\includegraphics[width=3in,clip]{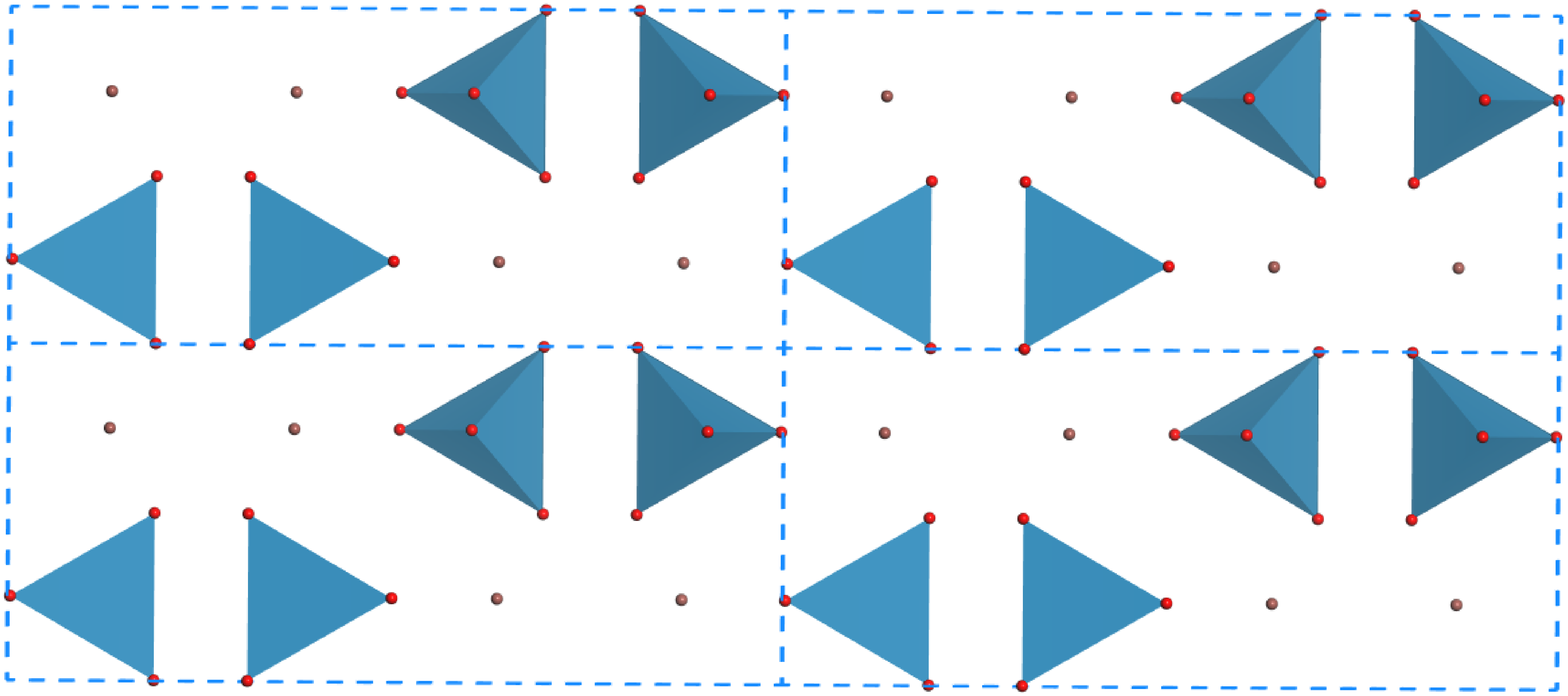}}
 \subfigure[]{\includegraphics[width=3in,clip]{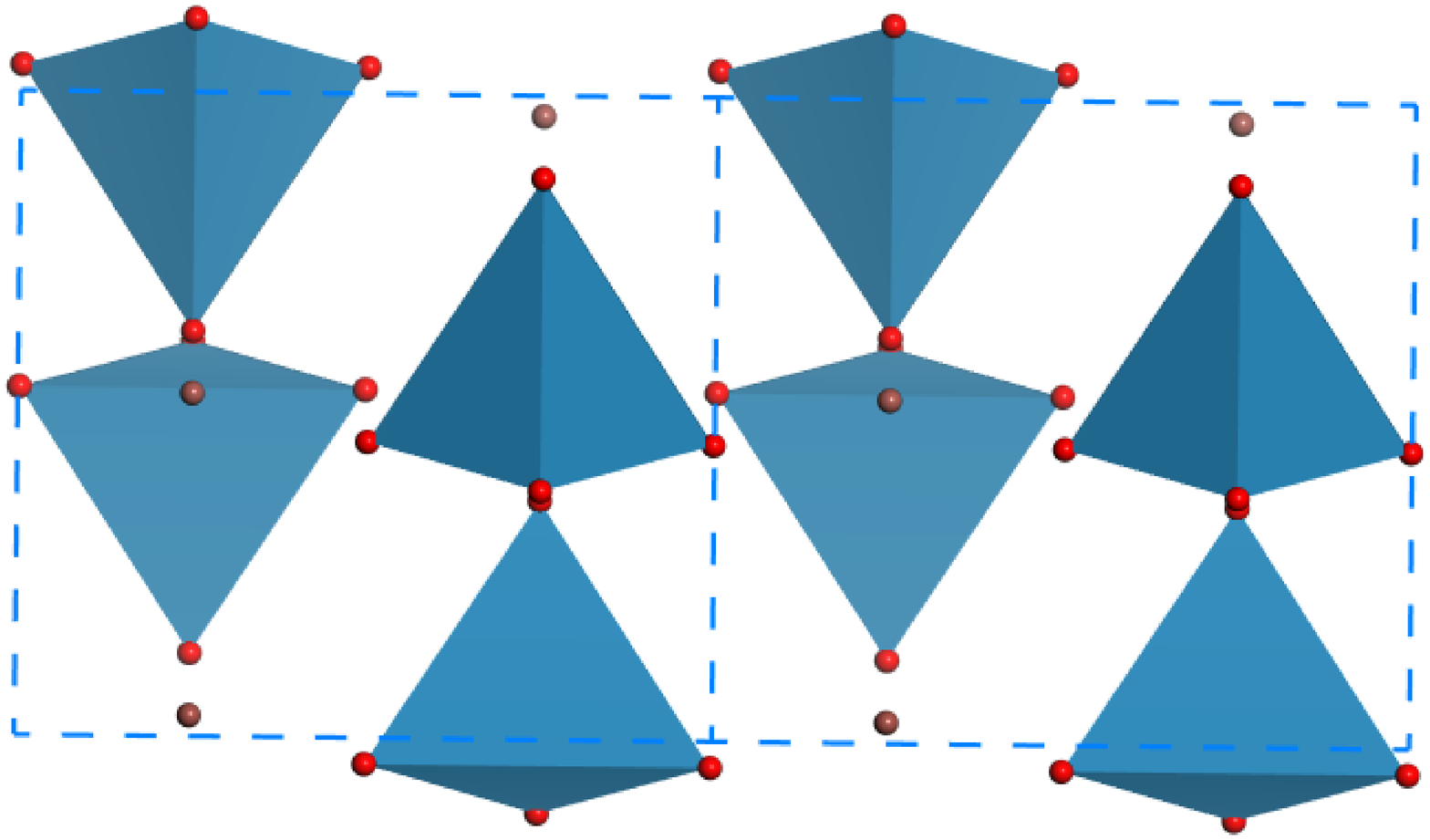}}
 \caption{(color online) Unit cell of orthorhombic TlReO$_4$ (shown in a) and super-cell of TlReO$_4$ oriented in yz-plane (shown in b) and
 xz-plane (shown in c).}
 \label{1}
\end{figure}

The ambient structure of TlReO$_4$ is the orthorhombic pseudoschelite structure with \textit{Pnma} space group where O3 type oxygen atoms can be found at 8d Wyckoff positions and Wyckoff positions of other atoms is 4c. A unit cell of orthorhombic TlReO$_4$ along with its super-cell oriented in yz and xy planes are shown in Fig \ref{1}. This figure clearly shows that the whole crystal is made up of ReO$_4$ tetrahedron building blocks and Tl atoms. Oxygen atoms of 8d Wyckoff positions in this tetrahedron are found to be oriented along the y-axis whereas other oxygen atoms can be found in the crystallographic xy-plane. It can also be seen that O2 type of oxygen atoms in this crystal are aligned along the x-axis and the interaction of these atoms with O1 type atoms of the nearest ReO$_4$ tetrahedron could favor the tilt or rotation of the tetrahedra upon the application of pressure. On the other hand, the oxygen atoms at 8d Wyckoff position could be the possible source for the suggested phase transition that happens in this material when pressure is applied.
In order to have a proper understanding of these speculative ideas drawn from the crystal structure, first principles calculations have been performed on this material with two different approaches based on Density Functional Theory (DFT)\cite{26,27}. A plane wave pseudo-potential approach (PW-PP) implemented in CASTEP code\cite{28,29} has been considered to investigate structural, elastic, and the vibrational properties of TlReO$_4$. An exchange-correlation functional characterized through local density approximation (LDA)\cite{30,31} and norm-conserving pseudo-potentials have been considered in this purpose. A cutoff energy of 975.0 eV has been used throughout the whole calculation and the Brillouin zone sampling was treated using Monkhorst-Pack scheme\cite{32} with k-points spacing 0.03 ${\AA}^{-1}$. For achieving the proper optimization of the crystal structure different convergence criteria, such as forces minimization criterion ($<$ 0.01 eV/$\AA$), energies minimization criterion ($<5 \times 10^{-6}$ eV), and stress tensors minimization criterion ($<$0.02 GPa) on each atom have been enforced. An extreme force minimization criterion of (1 $\times 10^{-4}$ eV/$\AA$) has also been applied in case of elastic constant calculation.\\
On the other hand, for calculating the electronic band structure and related optical properties of TlReO$_4$, a FP-LAPW method using WIEN2k package\cite{33} has been applied. The computationally less expensive TB-mBJ potential\cite{34} in addition to a standard GGA functional Perdew-Burke-Ernzerhof (PBE) has been considered to obtain a nearly exact band gap, as normal LDA and GGA functionals are known for underestimating the band gap in comparison to the experimental value. The presence of Re \emph{d}-states in this material also suggested to consider the inclusion of the spin-orbit coupling on top of the TB-mBJ calculation. All these calculations are done on the experimental crystal structure of TlReO$_4$\cite{7} and the number of K-points used for attaining the convergence is 500. An energy of -7.0 Ry has been used for separating the core and valence electrons in these calculations.

\section{Result and discussion}
\subsection{Structural and elastic properties}
Ground-state structural and mechanical properties of TlReO$_4$ have been discussed in this section. The experimental lattice parameters are considered as the initial input parameter in this regard. Different lattice parameters along with the bond lengths of this material have been calculated using local-density approximation (LDA) as well as the generalized gradient approximation (PBE). These parameters along with the atomic positions of the optimized crystal structure are tabulated in tables \ref{I} and \ref{II}.
\begin{table}
\caption{Calculated structural parameters of TlReO$_4$ at ambient condition.}
\begin{ruledtabular}
\begin{tabular}{cccc}
Parameters        & Expt.\footnote{Ref. \cite{7}}   & LDA     & PBE       \\
 a (in {\AA})     & 5.623   & 5.202   & 5.723     \\
 b (in {\AA})     & 5.791   & 5.658   & 5.965     \\
 c (in {\AA})     & 13.295  & 12.948  & 13.702    \\
 V (in {\AA}$^3$) & 432.922 & 381.073 & 467.793   \\
 Tl(4c)  & (0.02497   0.75000   0.12698) & (0.03678   0.75000   0.13313) & (0.03684   0.75000   0.13305) \\
 Re(4c)  & (0.03835   0.25000   0.37984) & (0.04095   0.25000   0.38009) & (0.01575   0.25000   0.38043) \\
 O1(4c) & (0.82200   0.25000   0.08380) & (0.87553   0.25000   0.09886) & (0.81606   0.25000   0.09050) \\
 O2(4c) & (0.85600   0.25000   0.47370) & (0.88272   0.25000   0.49970) & (0.85052   0.25000   0.48850) \\
 O3(8d) & (0.01220   0.51540   0.68590) & (0.04903   0.49720   0.69044) & (0.05534   0.50858   0.68803) \\
\end{tabular}
\end{ruledtabular}
\label{I}
\end{table}
Results shown in table \ref{I} follow the known trend of underestimation of lattice parameters with the LDA functional and overestimation of these parameters with the PBE functional. It is also noticeable from these data that value of lattice parameters 'b' and 'c' using LDA functional matches quite well with reported experimental results than corresponding values found from PBE functional whereas the lattice parameter 'a' is well reproduced with the PBE functional.
\begin{table}
\caption{Calculated bond lengths of TlReO$_4$.}
\begin{ruledtabular}
\begin{tabular}{cccc}
Bond length           & Expt.\footnote{Ref. \cite{7}}   & LDA     & PBE       \\
 Re$-$O1 (in {\AA})   & 1.667   & 1.762   & 1.764     \\
 Re$-$O2 (in {\AA})   & 1.615   & 1.754   & 1.757     \\
 Re$-$O3 ($\times2$)(in {\AA}) & 1.64    & 1.76    & 1.766     \\
 Tl$-$O1 (in {\AA})   & 2.931   & 3.038   & 3.177     \\
 Tl$-$O1 ($\times2$)(in {\AA}) & 3.165 & 2.984 & 3.291  \\
 Tl$-$O2 (in {\AA}) & 2.957 & 2.783 & 2.973 \\
 Tl$-$O3 ($\times2$)(in {\AA}) & 3.122 & 2.714 & 2.944  \\
 Tl$-$O3 ($\times2$)(in {\AA}) & 2.932 & 2.674 & 2.897 \\
\end{tabular}
\end{ruledtabular}
\label{II}
\end{table}
Though results shown in table \ref{I} clearly suggest to consider a van der Waals correction along with these exchange-correlation functionals, the different bond lengths tabulated in table \ref{II} give a different picture. All bond lengths with the LDA functional show nearly an equal value with the reported experimental bond lengths. The calculated distortion index (bond length) value of the ReO$_4$ tetrahedron with the LDA functional (0.0014) is found to be more regular than the PBE result (0.0018). In the case of TlO$_8$, the LDA result is near to the experimental result. Moreover, the bulk modulus value using LDA functional is found to match quite well with the experimental result which is reported to be 26 GPa\cite{19}. Therefore, a Van der Waals correction has not been included in this calculation. The bulk modulus along with its pressure derivative has been calculated using a 3rd-order Birch Murnaghan equation of state (BM-EOS)\cite{35} and the results are 29.77 GPa and 5.94 respectively. Considering all the above-mentioned points according to structural parameters, the LDA functional has been considered for the further calculations to obtain the mechanical and vibrational properties of TlReO$_4$.\\
To study the mechanical stability and the associated properties of this material, the LDA functional has been considered along with the optimized crystal structure as the next step. In this connection, the elastic constants of this material have been calculated using the stress-strain method. As the crystal structure is orthorhombic, we have found 9 elastic constant coefficients which are tabulated in table \ref{III} with other related mechanical parameters associated with these elastic constants. According to our knowledge, the elastic constants of TlReO$_4$ are being reported in this article for the first time. The mechanical stability of this crystal is confirmed from the born stability criterion. The order of the elastic constant along the principal axis, C$_{22}$ $>$ C$_{33}$ $>$ C$_{11}$ tells the anisotropic nature of this crystal. As discussed in the crystal structure and computational details section, the presence of O3 oxygen atoms along the y-axis and O1 and O2 types of oxygen atoms in the xz-plane could be the possible origin of this anisotropy. The poly-crystalline bulk modulus found from the elastic constant calculation is reported to be 36.46 GPa which is slightly higher than the bulk modulus found from BM-EOS. The bulk modulus value of TlReO$_4$ is found to be much smaller compared to other ABO$_4$-type scheelite materials CaWO$_4$ (74 GPa), SrWO$_4$ (63 GPa)\cite{38}, CaMoO$_4$ (82 GPa)\cite{39} and SrMoO$_4$ (71 GPa)\cite{40}. This is because of the large volume of the TlO$_8$ dodecahedron which dominates the compressibility of the crystal and is very compressible due to the weak Tl-O bonds (which is a consequence of the electronic distribution, highly localized around Re and delocalized around Tl).
\begin{table}
\caption{Computed elastic constant values C$_{ij}$ and associated parameters such as: Bulk modulus (B), isothermal compressibility (K) and young's modulus (E). Units are as follows: C$_{ij}$ in GPa, B in GPa, K in GPa$^{-1}$ and E in GPa.}
\begin{ruledtabular}
\begin{tabular}{cccccccccccc}
C$_{11}$&C$_{12}$&C$_{13}$&C$_{22}$&C$_{23}$&C$_{33}$&C$_{44}$&C$_{55}$&C$_{66}$&  B  &   K & E  \\
 53.50  & 29.80  & 26.17  & 61.71  & 22.70  & 56.65  & 13.77  & 19.82  & 16.88  &36.46&0.027& 34 \\
\end{tabular}
\end{ruledtabular}
\label{III}
\end{table}

\subsection{Born effective charge (BEC)}
To have a better understanding of the material properties such as vibrational and optical properties, the born effective charge is a very useful fundamental quantity. The quantity itself defines the coupling of the electrostatic field with the lattice displacement. In this study, the linear response method implemented in CASTEP has been used for the purpose of computing BEC. All the related results are shown in table \ref{IV}. The convergence of this calculation satisfies the acoustic sum rule which is $\sum_kZ^*_{k,ii}=0$. The result in table \ref{IV} clearly shows that all the components of BEC are not null for O3 atoms whereas the other atoms present in this material have only diagonal elements along with Z$_{13}^*$ and Z$_{31}^*$ components of BECs. This clarifies the absence of interaction in xy and yz plane due to Tl, Re, O1 and O2 atoms. Though a nearly doubling of nominal charge for Tl and a large reduction of nominal charge of Re atoms in xx, yy and zz components of BECs can be seen but it follows a Z$_{11}^* \neq $Z$_{22}^* \neq $Z$_{33}^*$ relation for both elements. Among oxygen atoms, O1 atoms are found to follow a different trend than other oxygens.\\
The existence of a mix covalent-ionic nature of bonding is quite evident from the deviation of BEC's with respect to the minimal charge of the different elements present in this material. The covalent nature of any bond can be seen as transfer of charge in a considerable amount upon displacement of the atoms associated to that bond\cite{36,37}. In this connection, the large deviation of the nominal charge of all the elements present in TlReO$_4$ helps us to make a conclusion that this material possesses a notable covalent character. Though a large reduction of the nominal charge of Re and oxygen atoms in ReO$_4$ anions can be seen, a smaller reduction of the nominal charge of O1 oxygens introduces ionicity in this covalent building block. Moreover, the presence of off-diagonal elements for O3 oxygens supports the earlier comment. Along with this, the comparatively large amount of BEC of the Tl atom indicates a possible hybridization of Tl and neighboring O1 oxygens.
\begin{table}
\caption{Calculated BECs of Tl, Re and O atoms of TlReO$_4$.}
\begin{ruledtabular}
\begin{tabular}{ccccccccccc}
Atoms & Ionic & Z$_{11}^*$ & Z$_{12}^*$ & Z$_{13}^*$ & Z$_{21}^*$ & Z$_{22}^*$ & Z$_{23}^*$ & Z$_{31}^*$ & Z$_{32}^*$ & Z$_{33}^*$ \\
Tl1   &   1   &  2.006    &    0       &  0.209   &     0      & 2.166    &      0     &   0.270  &     0      &  2.077   \\
Tl2   &   1   &  2.006    &    0       & -0.209   &     0      & 2.166    &      0     &  -0.270  &     0      &  2.077   \\
Re1   &   7   &  3.117    &    0       &  0.184   &     0      & 3.296    &      0     &   0.442  &     0      &  3.213   \\
Re2   &   7   &  3.117    &    0       & -0.184   &     0      & 3.296    &      0     &  -0.442  &     0      &  3.213   \\
O1    &  -2   & -2.529   &    0       &  0.334   &     0      & -0.774    &      0     &   0.381  &     0      & -0.724   \\
O1    &  -2   & -2.529   &    0       & -0.334   &     0      & -0.774    &      0     &  -0.381  &     0      & -0.724   \\
O2    &  -2   & -0.914   &    0       &  0.921   &     0      & -0.494   &      0     &   0.747  &     0      & -1.954   \\
O2    &  -2   & -0.914   &    0       & -0.921   &     0      & -0.494   &      0     &  -0.747  &     0      & -1.954   \\
O3    &  -2   & -0.839   &  0.713   & -0.195   &  0.527   & -2.096   &  0.883   &  -0.191  &  1.041   & -1.305   \\
O3    &  -2   & -0.839   &  0.713   &  0.195   &  0.527   & -2.096   & -0.883   &   0.191  & -1.041   & -1.305   \\
O3    &  -2   & -0.839   & -0.713   & -0.195   & -0.527   & -2.096   & -0.883   &  -0.191  &  1.041   & -1.305   \\
O3    &  -2   & -0.839   & -0.713   &  0.195   & -0.527   & -2.096   &  0.883   &   0.191  & -1.041   & -1.305   \\
\end{tabular}
\end{ruledtabular}
\label{IV}
\end{table}

\subsection{Vibrational properties}
The optimized crystal structure with the LDA functional has been used for calculating zone centered phonon modes and associated IR spectra for this material. In this purpose, linear response theory implemented in the density functional perturbation theory (DFPT) is considered. As the primitive cell of TlReO$_4$ is constituted by 24 atoms, 72 phonon modes are expected at $\Gamma$-point. Among them, 3 modes are acoustic modes whereas the rest 69 modes are optical modes. The optical modes are sub-divided into 3 types depending on symmetry: IR-active modes (B1$_u$, B2$_u$, B3$_u$), Raman-active modes (A$_g$, B1$_g$, B2$_g$, B3$_g$) and silent modes (A$_u$). Among them, 33 modes are also called external modes as these modes are due to whole lattice vibrations and 36 modes are called internal modes as these modes are due to the stretching and bending of the ReO$_4$ tetrahedron. The irreducible representation of these phonon modes can be written as \\
$\Gamma = \Gamma_{acoustic} + \Gamma_{IR} + \Gamma_{Raman} + \Gamma_{silent}$ \\
Here,\\
 $\Gamma_{acoustic} = B2_{u} + B1_{u} + B3_{u}$\\
$\Gamma_{IR} = 10B1_u + 6B2_u + 10B3_u$ \\
$\Gamma_{Raman} = 11 A_g + 7 B1_g + 11B2_g + 7 B3_g$\\
$\Gamma_{silent} = 7 A_u$ \\
All these modes are non-degenerate.
\begin{table}
\caption{Calculated zone centered vibrational modes of TlReO$_4$ from 0 to 200 cm$^{-1}$ along with experimental value (see in Ref. \onlinecite{22}) and assignment of different modes. [M1-M3 are acoustic modes, Rot.=Rotation, Trans.=Translational, Rock.=Rocking, Wagg.=Wagging and Twist.=Twisting]}
\scalebox{0.8}{
\begin{ruledtabular}
\begin{tabular}{cccccc}
Mode No. & Frequency (cm$^{-1}$) & Expt (cm$^{-1}$)  & Symmetry & Mode    & Assignment      \\ \hline
M4       &  6.90   	         &                   & B2$_u$      &  	IR    & Rot. of ReO$_4$      \\
M5	     &  10.68   	     &                   & B1$_g$      &  	Raman &	Trans. of (Tl, ReO$_4$)   \\
M6       &	26.05            &                   & A$_u$       &  Silent &	Trans. of (Tl, ReO$_4$)   \\
M7       &	28.28   	     &                   & A$_g$       &  	Raman &	Trans. of (Tl, ReO$_4$)  \\
M8       &	32.37   	     &                   & B3$_g$      &  	Raman &	Trans. of (Tl, ReO$_4$)   \\
M9       &	34.29            &                   & B1$_u$      &  	IR	  &	Trans. of (Tl, ReO$_4$)  \\
M10      &	36.04   	     &                   & B3$_u$      &  	IR	  &	Rot. of ReO$_4$, Trans. of Tl     \\
M11      &	43.51   	     &  39               & B2$_g$      &  	Raman &	Trans. of (Tl, ReO$_4$)   \\
M12      &	44.03            &                   & A$_u$       &  Silent &	Trans. of (Tl, ReO$_4$)   \\
M13	     &  45.48   	     &                   & A$_g$       & 	Raman &	Trans. of (Tl, ReO$_4$)   \\
M14	     &  45.81   	     &                   & B1$_u$      &  	IR	  &	Trans. of (Tl, ReO$_4$)  \\
M15	     &  48.92   	     &  49               & B2$_g$      &  	Raman & Trans. of (Tl, ReO$_4$)   \\
M16	     &  51.20   	     &                   & B1$_g$      &  	Raman &	Rot. of ReO$_4$      \\
M17	     &  54.91   	     &                   & B3$_g$      &  	Raman &	Trans. of (Tl, ReO$_4$)   \\
M18	     &  60.11   	     &                   & B3$_u$      &  	IR	  &	Trans. of (Tl, ReO$_4$)   \\
M19	     &  60.68   	     &                   & B2$_u$      &  	IR	  &	Trans. of (Tl, ReO$_4$)   \\
M20	     &  60.94   	     &  62               & A$_g$       &  	Raman &	Trans. of (Tl, ReO$_4$)   \\
M21      &	63.41   	     &                   & B1$_u$      &  	IR	  &	Trans. of (Tl, ReO$_4$)   \\
M22	     &  70.78            &                   & B3$_u$      &  	IR	  &	Trans. of (Tl, ReO$_4$)   \\
M23	     &  72.10   	     &                   & B2$_g$      &  	Raman & Trans. of (Tl, ReO$_4$)   \\
M24	     &  73.59   	     &                   & B1$_g$      &  	Raman & Rot. of ReO$_4$     \\
M25	     &  76.83   	     &                   & B2$_g$      &  	Raman & Trans. of (Tl, ReO$_4$)   \\
M26	     &  77.75   	     &                   & A$_g$       &  	Raman & Trans. of ReO$_4$   \\
M27	     &  84.21   	     &                   & B3$_g$      &  	Raman &	Rot. of ReO$_4$      \\
M28	     &  85.59   	     &                   & A$_u$       &  Silent & Rot. of ReO$_4$      \\
M29	     &  93.74   	     &                   & A$_u$       &  Silent & Rot. of ReO$_4$  \\
M30	     &  101.70   	     &                   & B3$_u$      &  	IR	  &	Rot. of ReO$_4$   \\
M31	     &  116.94   	     &                   & B3$_g$      &  	Raman &	Rot. of ReO$_4$, Twist. in ReO$_4$     \\
M32	     &  117.59   	     &                   & A$_g$       & 	Raman & Rot. of ReO$_4$      \\
M33	     &  119.28   	     &                   & B2$_u$      &  	IR	  &	Rock., Wagg. in ReO$_4$      \\
M34	     &  124.22   	     &                   & B2$_g$      &  	Raman &	Rot. of ReO$_4$      \\
M35	     &  127.29   	     &                   & B1$_g$      &  	Raman & Rock., Wagg. in ReO$_4$      \\
M36	     &  135.34   	     &                   & B1$_u$      &  	IR	  & Rot. of ReO$_4$      \\
\end{tabular}
\end{ruledtabular}}
\label{V}
\end{table}

\begin{table}
\caption{Calculated zone centered vibrational modes of TlReO$_4$ in the intermediate frequency range with experimental value (see in Ref.
 \onlinecite{22}) and associated mode assignment. [Rock.=Rocking, Wagg.=Wagging, Sci.=Scissoring,  Twist.=Twisting]}
\begin{ruledtabular}
\begin{tabular}{cccccc}
Mode No.  &  Frequency (cm$^{-1}$) & Expt (cm$^{-1}$) & Symmetry  & Mode    & Assignment  \\ \hline
M37	      &  266.76   	       &                  & B2$_u$       & IR	    & Rock., Wagg., Sci. in ReO$_4$    \\
M38	      &  267.89   	       &                  & A$_u$        & Silent   & Rock., Wagg., Sci. in ReO$_4$    \\
M39	      &  296.42   	       &                  & A$_g$        & Raman	& Wagg., Sci. in ReO$_4$    \\
M40	      &  296.44   	       &                  & B3$_u$       & IR	    & Wagg., Sci. in ReO$_4$    \\
M41	      &  297.47   	       &                  & B1$_g$       & Raman	& Sci., Twist. in ReO$_4$    \\
M42	      &  299.10   	       &                  & B1$_u$       & IR	    & Sci. in ReO$_4$    \\
M43	      &  307.84   	       &                  & B2$_g$       & Raman	& Sci., Wagg. in ReO$_4$    \\
M44	      &  310.26   	       &                  & B3$_g$       & Raman	& Sci., Twist. in ReO$_4$    \\
M45	      &  313.25   	       &                  & B2$_u$       & IR	    & Twist., Sci. in ReO$_4$  \\
M46	      &  313.52   	       &                  & A$_g$        & Raman	& Sci. in ReO$_4$    \\
M47	      &  314.65   	       &                  & B3$_u$       & IR	    & Sci. in ReO$_4$    \\
M48	      &  316.68   	       &                  & B1$_u$       & IR	    & Wagg., Sci. in ReO$_4$   \\
M49	      &  322.21   	       &                  & A$_u$        & Silent   & Twist., Sci. in ReO$_4$   \\
M50	      &  323.50   	       &    324           & B2$_g$       & Raman	& Wagg., Sci. in ReO$_4$    \\
M51	      &  326.48   	       &                  & A$_g$        & Raman	& Wagg., Sci. in ReO$_4$      \\
M52	      &  326.48   	       &    329           & B2$_g$       & Raman	& Sci. in ReO$_4$     \\
M53	      &  335.60   	       &    335           & B3$_g$       & Raman	& Twist., Sci. in ReO$_4$    \\
M54	      &  337.74   	       &    339           & B1$_g$       & Raman	& Sci., Twist. in ReO$_4$    \\
M55	      &  338.65   	       &                  & B3$_u$       & IR	    & Sci., Twist. in ReO$_4$    \\
M56	      &  339.54   	       &                  & B1$_u$       & IR	    & Sci., Twist. in ReO$_4$      \\
\end{tabular}
\end{ruledtabular}
\label{VI}
\end{table}

\begin{figure}
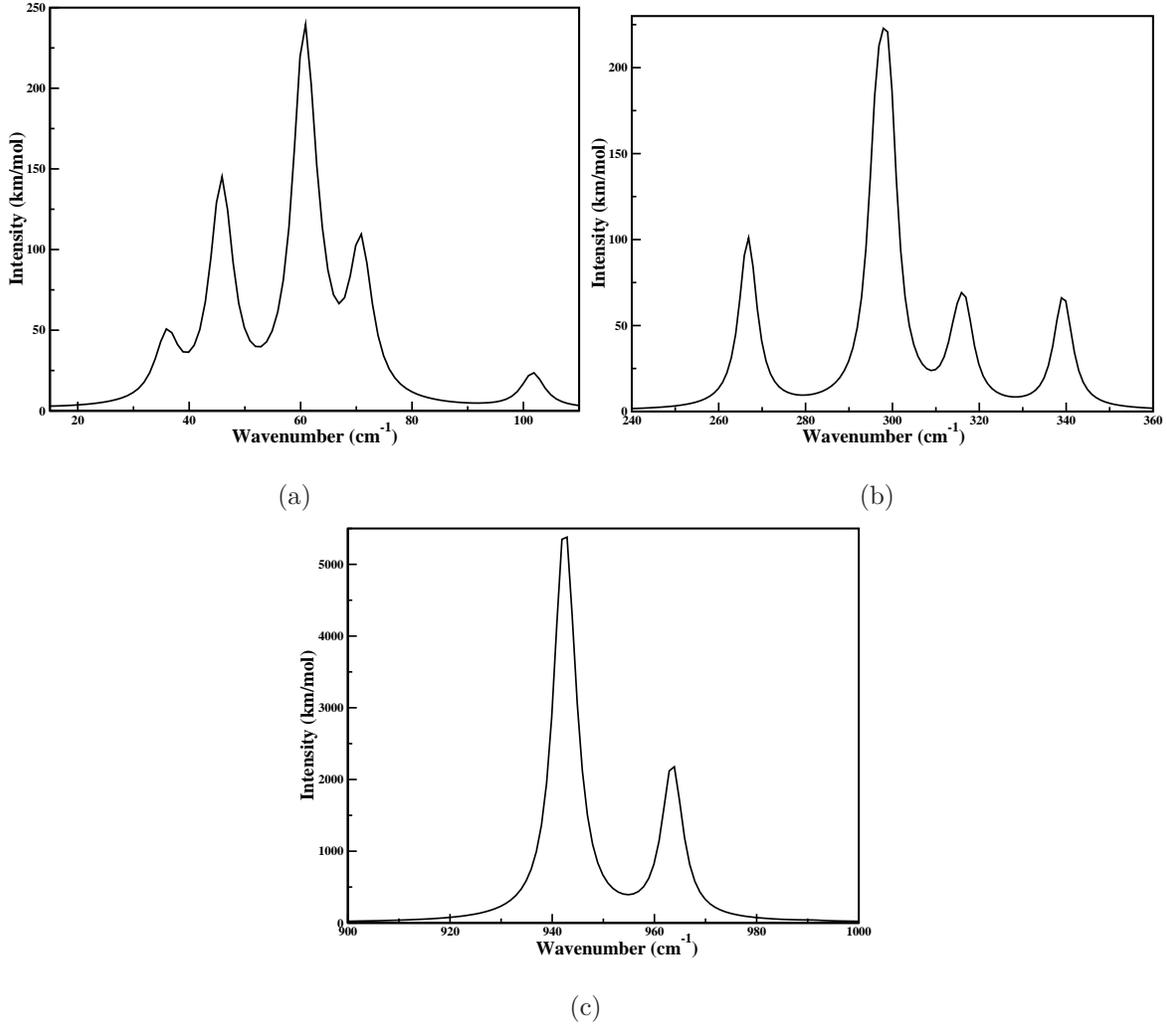

 \subfigure[]{\includegraphics[width=3in,clip]{4.eps}}
 \subfigure[]{\includegraphics[width=3in,clip]{5.eps}}
 \subfigure[]{\includegraphics[width=3in,clip]{6.eps}}
 \caption{(color online) IR spectra of TlReO$_4$ in three different regions.}
 \label{2}
\end{figure}

\begin{table}
\caption{Calculated zone centered vibrational modes of TlReO$_4$ higher than 900 cm$^{-1}$ with experimental value (see in Ref. \onlinecite{22}). [Asym.=Asymmetric, Sym.=Symmetric, Str.=Stretching]}
\begin{ruledtabular}
\begin{tabular}{cccccc}
Mode No.  & Frequency (cm$^{-1}$) & Expt (cm$^{-1}$) & Symmetry & Mode   & Assignment  \\ \hline
M57	      & 930.83   	      &                  & A$_u$       & Silent  & Asym. Str. of Re-O3 bonds   \\
M58	      & 931.71    	      &                  & B3$_g$      & Raman   & Asym. Str. of Re-O3 bonds  \\
M59	      & 941.92   	      &                  & A$_g$       & Raman   & Asym. Str. of Re-O1,\\
          &                   &                  &             &         & sym. str. of Re-O3   \\
M60	      & 942.07   	      &                  & B2$_u$      & IR	     & Asym. Str. of Re-O3 bonds   \\
M61	      & 942.80   	      &                  & B3$_u$      & IR	     & Str. of Re-O1 bond     \\
M62	      & 943.13   	      &                  & B1$_g$      & Raman   & Asym. Str. of Re-O3 bond   \\
M63	      & 944.74   	      &                  & B1$_u$      & IR	     & Str. of Re-O1 bond   \\
M64	      & 951.02   	      &                  & B2$_g$      & Raman   & Asym. Str. of Re-O1,\\
          &                   &                  &             &         & Re-O2 bonds   \\
M65	      & 960.11   	      &   959            & A$_g$       & Raman   & Str. of Re-O2 bond    \\
M66	      & 963.50   	      &                  & B1$_u$      & IR	     & Sym. Str. of Re-O3, \\
          &                   &                  &             &         & Asym. Str. of Re-O2 \\
M67	      & 968.88   	      &                  & B2$_g$      & Raman   & Sym. Str. of Re-O3, \\
          &                   &                  &             &         & Asym. Str. of Re-O2  \\
M68	      & 991.01   	      &                  & B3$_u$      & IR	     & Sym. Str. of Re-O3, \\
          &                   &                  &             &         & Asym. Str. of Re-O2  \\
M69	      & 1013.57   	      &                  & A$_g$       & Raman   & Sym. Str. of all bonds in ReO$_4$    \\
M70	      & 1013.90   	      &                  & B2$_g$      & Raman   & Sym. Str. of all bonds in ReO$_4$    \\
M71	      & 1016.33   	      &                  & B1$_u$      & IR	     & Sym. Str. of all bonds in ReO$_4$    \\
M72	      & 1018.44   	      &                  & B3$_u$      & IR	     & Sym. Str. of all bonds in ReO$_4$    \\
\end{tabular}
\end{ruledtabular}
\label{VII}
\end{table}
All zone centered phonon frequencies along with the assigned symmetries are tabulated in Tables \ref{V}-\ref{VII}. The complete IR spectra (including intensities) of this material has also been calculated and shown in figure \ref{2}. In this study, the whole IR spectra is divided into three regions for a better resolution.
These regions are as follows: Frequencies attributed to external lattice vibrations or the mixture of lower molecular vibrations which is from 0 to 200 cm$^{-1}$, Frequencies from 240 cm$^{-1}$ to 360 cm$^{-1}$ and the frequencies above 900 cm$^{-1}$.
\begin{figure}
 \subfigure[]{\includegraphics[width=2in,clip]{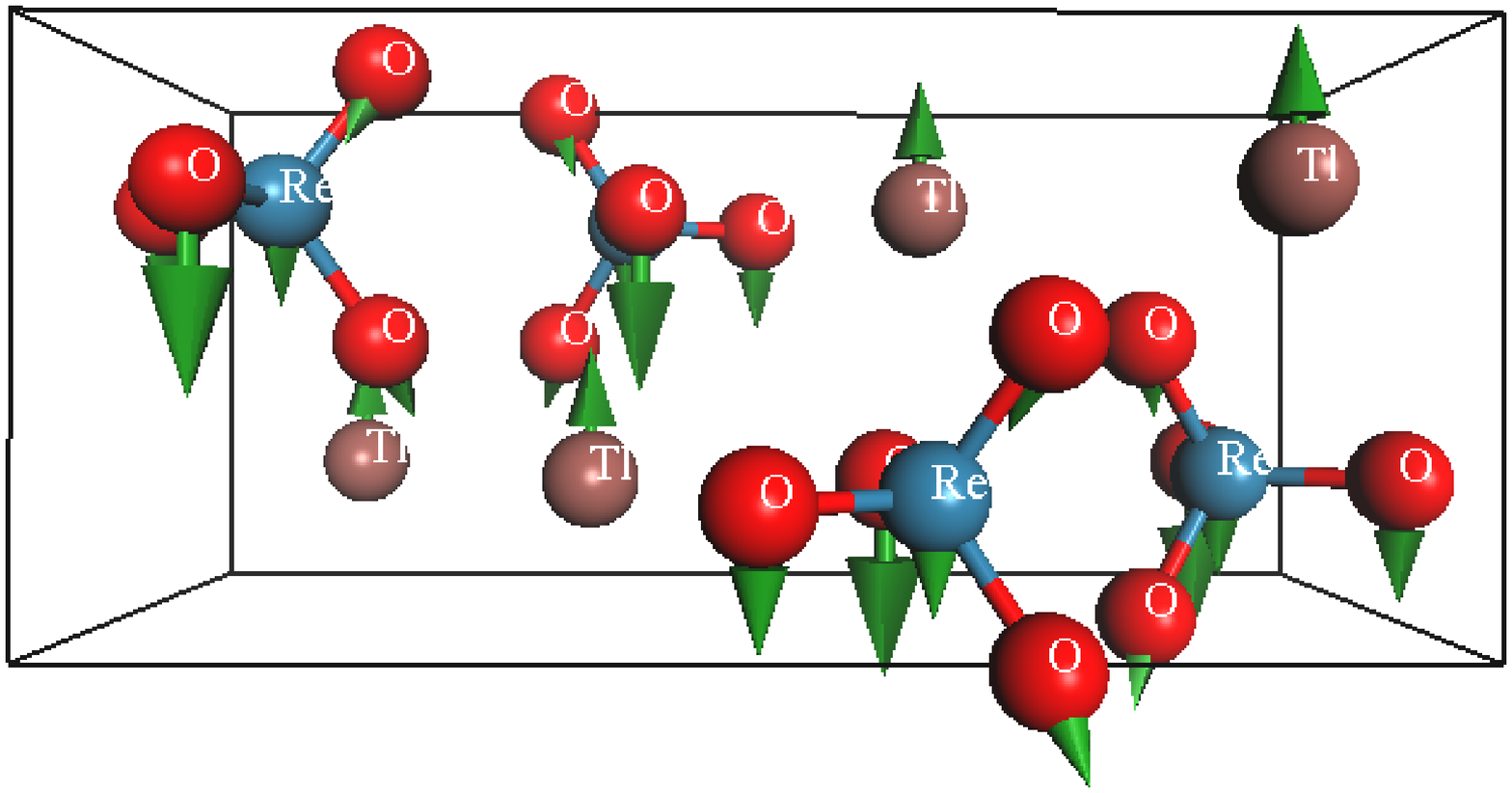}}
 \subfigure[]{\includegraphics[width=2in,clip]{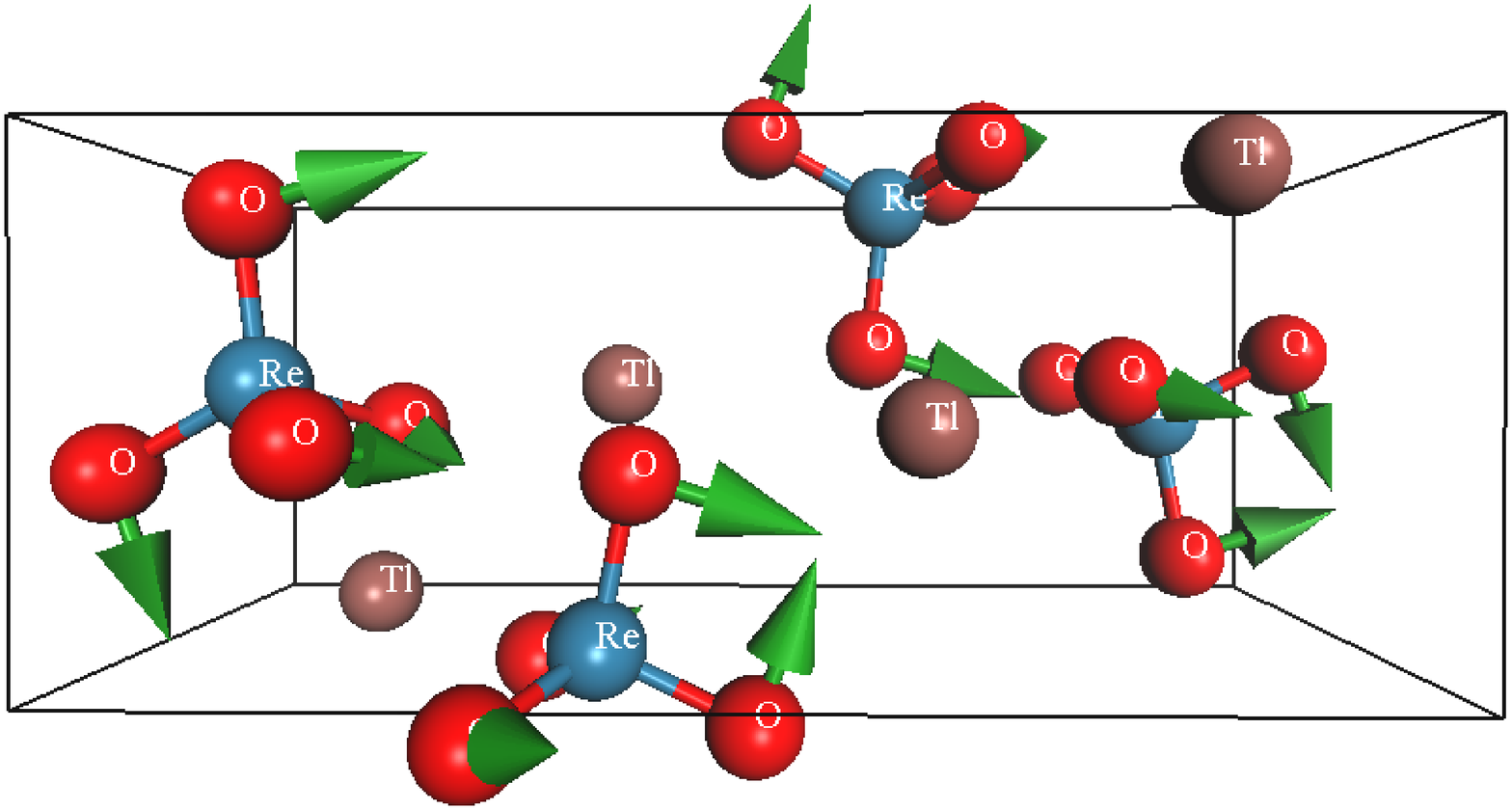}}
 \subfigure[]{\includegraphics[width=2in,clip]{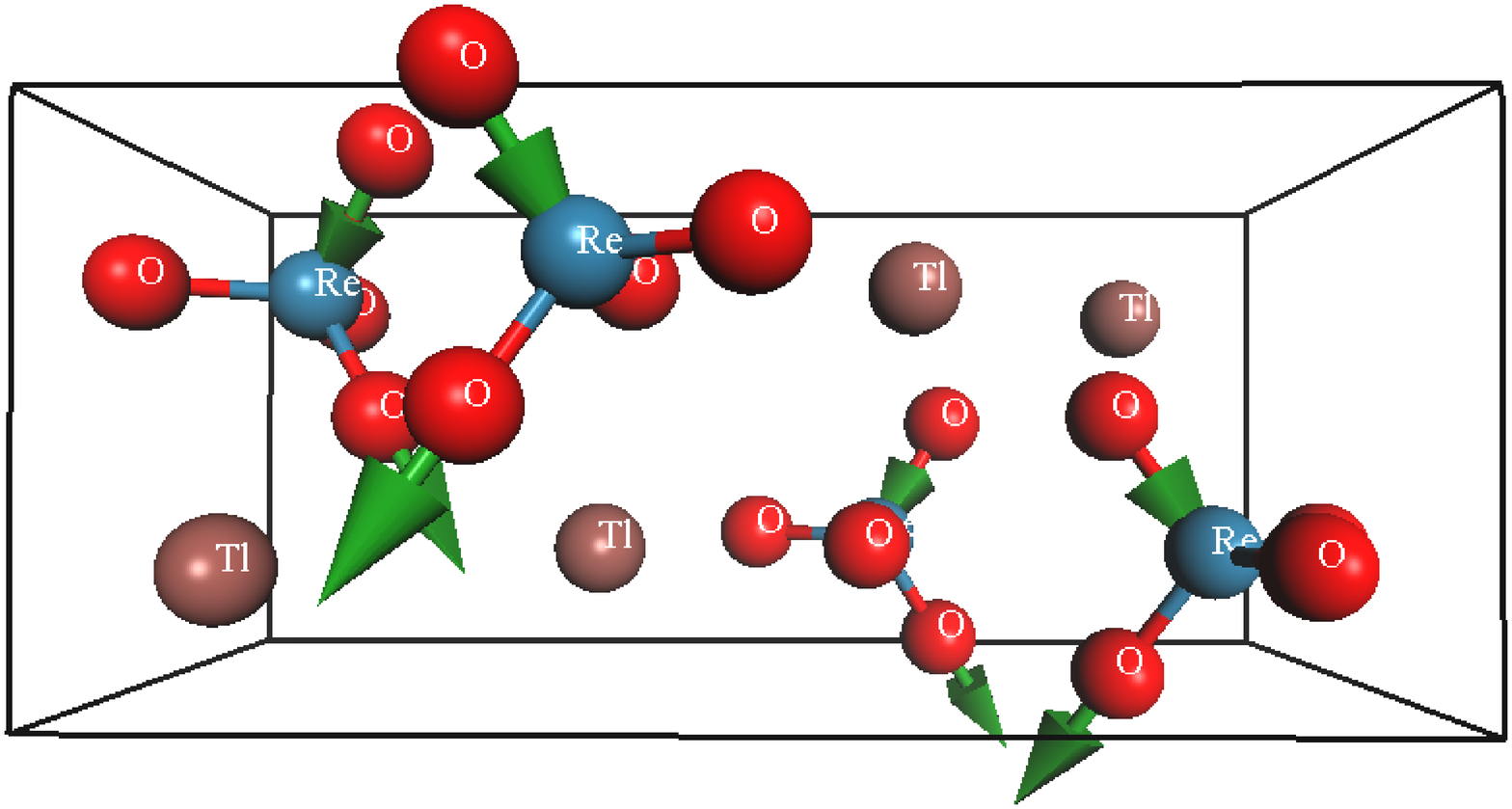}}
 \caption{(color online) Assigned vibrational modes of most intense IR peak in different range.}
 \label{3}
\end{figure}
In order to explain the first frequency range, the IR spectra of TlReO$_4$ from 15 to 110 cm$^{-1}$ has been depicted in figure \ref{2}(a). This figure clearly tells that 4 very intense peaks are present in this spectra. These peaks belong to the external lattice vibrations. Among them, the highest intensity peak is found at 60.74 cm$^{-1}$ and the associated vibrational is shown in figure \ref{3}(a). Figure \ref{3}(a) clearly shows a collective vibration of Tl atoms and ReO$_4$ groups in the opposite direction which is the main reason behind the intense peak at 60.74 cm$^{-1}$. The second highest intense peak is found at 45.93 cm$^{-1}$. This mode is due to a similar kind of translation of Tl atoms and ReO$_4$ groups. The two peaks that follow in intensity are at 70.82 and at 36.14 cm$^{-1}$ being associated with a translation of Tl and ReO$_4$. The IR spectra of TlReO$_4$ in the 240 to 360 cm$^{-1}$ frequency range also shows 4 prominent peaks which can be seen in figure \ref{2}(b). The most intense peak is due to the scissoring of the ReO$_4$ group and the frequency is 299.15 cm$^{-1}$. The corresponding vibration can be seen in the figure \ref{3}(b). Other intense peaks which are also due to a similar scissoring of ReO$_4$ group are found at 266.79, 316.68 and 339.54 cm$^{-1}$. In case of high-frequency range above 900 cm$^{-1}$, only two prominent peaks can be seen which are shown in figure \ref{2}(c). Among these peaks, the strongest peak is located at 942.07 cm$^{-1}$ and the second most intense mode is at 963.5 cm$^{-1}$. An asymmetric stretching of Re$-$O3 bonds in ReO$_4$ group is found to be the reason for the first one whereas a symmetric stretching of Re$-$O3 bonds along with asymmetric stretching of Re$-$O2 bonds is the reason for the other. The movements associated to the strongest peak are shown in figure \ref{3}(c).

\subsection{Electronic band structure}
\begin{figure}
 \centering
 \includegraphics[width=4in]{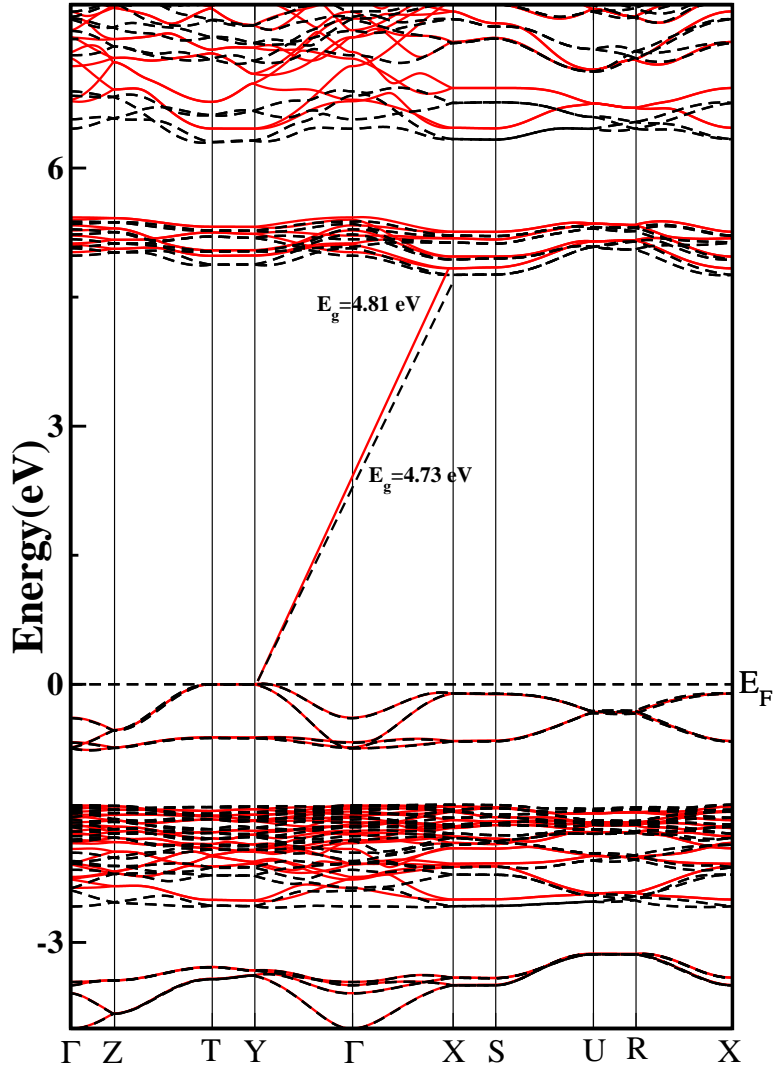}\\
 \caption{(color online) Electronic band structure of TlReO$_4$ with TB-mBJ potential (red) as well as TB-mBJ+SO scheme (black).}
 \label{4}
\end{figure}

\begin{figure}
 \centering
 \includegraphics[width=4in]{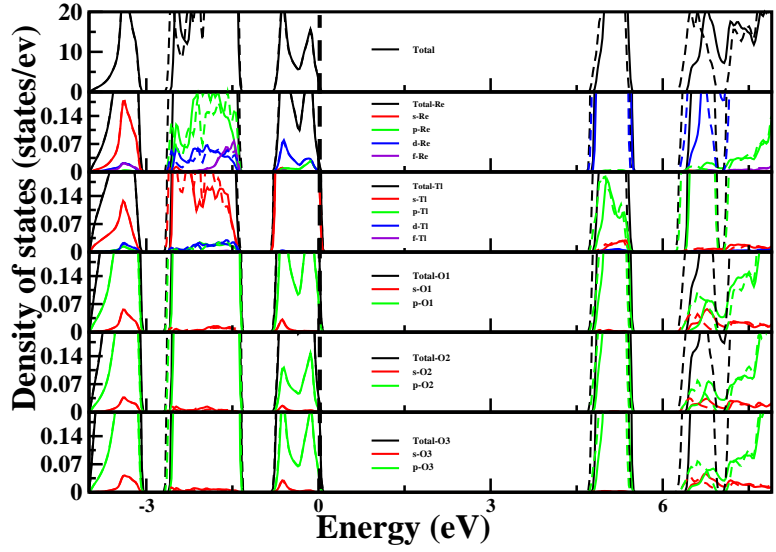}\\
 \caption{(color online) Density of states of TlReO$_4$ with TB-mBJ potential as well as TB-mBJ+SO scheme.}
 \label{5}
\end{figure}

A calculation of the electronic band structure of orthorhombic TlReO$_4$ has been carried out using the FP-LAPW approach. The experimental lattice parameters have been considered as the input parameters in this regard and the electronic band structure of TlReO$_4$ is depicted in figure \ref{4}. In order to obtain the exact band gap, a semi-local
potential TB-mBJ along with the spin-orbit coupling (SOC) have also been considered. The computed band gap using the LDA functional is found to be 2.4 eV whereas the band gap value has increased to 4.81 eV after the inclusion of TB-mBJ potential. The inclusion of SOC in the band structure calculation has decreased the band gap to 4.73 eV. Figure \ref{4} clarifies the nature of this band structure which is indirect with conduction-band bottom at X and valence-band top at Y high symmetric direction of the Brillouin zone. Our calculated band structure explains the transmission reported by Kunkley \cite{41}. They observed a sharp change in the transmission at 280 nm (4.4 eV) plus a lower energy exponential decay in transmission which resembles an Urbach tail which extends below the fundamental band gap. As it is not trivial to understand the contribution of different states of particular atom from the band structure, the density of states as well as the partial density of states have also been evaluated and plotted in figure \ref{5}. It clearly illustrates that the density of states (plotted from -4 eV to 8 eV) at energies lower than the Fermi level is divided into 3 regions whereas
\begin{figure}
 \subfigure[]{\includegraphics[width=4in,clip]{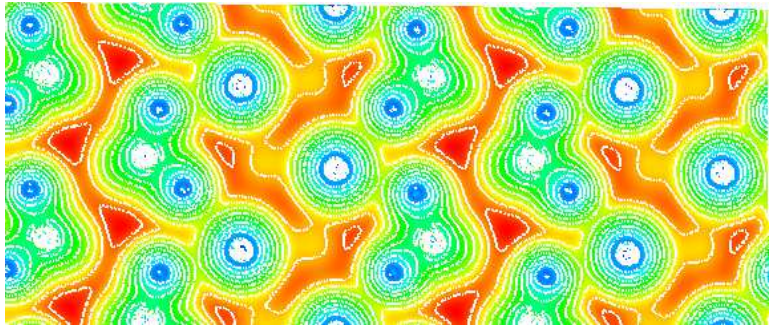}}
 \subfigure[]{\includegraphics[width=4in,clip]{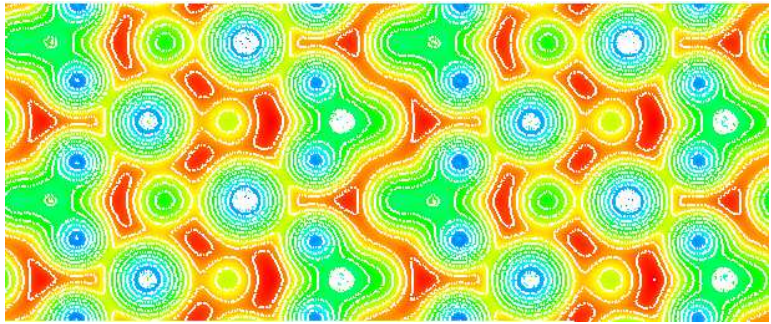}}
 \caption{(color online) 2D electronic charge density difference plot of TlReO$_4$ projected in (a) (101) plane and (b) (011) plane.}
 \label{6}
\end{figure}
the conduction band is divided into 2 regions. The contributions to the band structure from -4 eV to -3 eV are mainly due to the oxygen \emph{p}-states and Tl and Re \emph{s}-states. The middle region of the valence band which is from -2.6 eV to -1.35 eV is mainly made up of oxygen \emph{p}-states, Tl \emph{s}-states and Re \emph{p}-states along with a small contribution of \emph{d}-states. The valence band near the Fermi level is due to oxygen \emph{p}-states, Tl \emph{s}-states and a slight contribution from Re \emph{d}-states. The conduction band up to 8 eV shows also two regions where the region below 6 eV is mainly made of Re \emph{d}-states and oxygen \emph{p}-states. The region above 6 eV is dominated by Re \emph{d}-states, Tl \emph{p}-states and a small contribution of oxygen \emph{p}-states. This analysis of the density of states clearly shows the covalent nature of bonding between oxygen and Re atoms and the ionic nature between Tl and the ReO$_4$ anion. To validate this, the charge density mapping has been evaluated in (101) and (011) planes as the xz-plane can be considered for O1 and O2 oxygens whereas the O3 oxygens can be found in the yz-plane. The calculated charge density distribution is shown in Fig. \ref{6}. It clearly shows a considerable sharing of charges between Re and oxygen atoms and a small sharing of charges among Tl and other atoms. It can also be seen from the charge density plot in the yz-plane that the Tl and oxygen atoms are having ionic bond whereas the charge density plot in xz-plane clearly shows a little sharing of charge between Tl and O1 oxygens. This result is found to be consistent with the conclusions drawn from the DOS.

\subsection{Optical properties}
The calculation of the optical properties of TlReO$_4$ has been done using the local potential TB-mBJ with a SOC interaction. The absorption coefficient and refractive index of this material have been calculated using the dielectric function. The imaginary and
\begin{figure}
 \centering
 \includegraphics[width=4in]{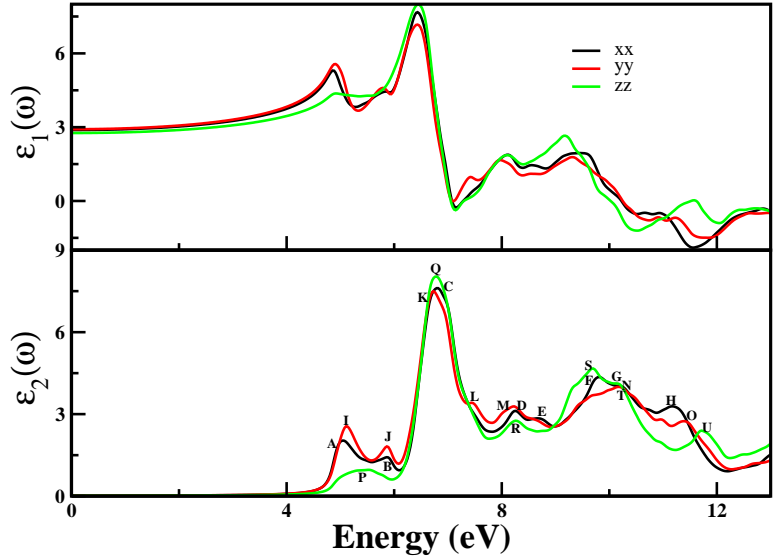}\\
 \caption{(color online) Calculated complex dielectric function of TlReO$_4$ using experimental crystal structure. $\varepsilon_1(\omega)$ indicating real part and the $\varepsilon_2(\omega)$ indicating imaginary part of this function are depicted in up and below panel, respectively.}
 \label{7}
\end{figure}
real dielectric function of this material are depicted in figure \ref{7}. The prominent peaks along \textit{XX} direction can easily be seen at A, B, C, D, E, F, G, H points of this figure whereas peaks along \textit{YY} direction can be found out at I, J, K, L, M, N, O points and along \textit{ZZ} direction at P, Q, R, S, T, U. It is well known that these peaks in the imaginary dielectric functions appear due to the inter-band transitions from the conduction band to the valence band. These peaks of the imaginary part of dielectric function along with the electronic band structure have been used for calculating a possible inter-band transitions of this material. The calculated transitions are found as: (a) O(\emph{p}-states)$\rightarrow$Re(\emph{d}-states) in between 4.8 to 6.2 eV, (b) Tl(\emph{s}-states)$\rightarrow$O(\emph{p}-states) in between 4.8 to 6.2 eV, (c) O(\emph{p}-states)$\rightarrow$Re(\emph{d}-states) in between 6.2 to 8.2 eV, (d) Re(\emph{d}-states)$\rightarrow$O(\emph{p}-states) in between 6.2 to 8.2 eV, (e) Tl(\emph{s}-states)$\rightarrow$O(\emph{p}-states) in between 6.2 to 8.2 eV, (f) O(\emph{p}-states)$\rightarrow$Re(\emph{d}-states) in between 8.2 to 9.2 eV. The important transition
\begin{table}[t]
\caption{Possible optical transitions of TlReO$_4$ using imaginary part of dielectric function and electronic band structure.}
\begin{center}
{\setlength{\tabcolsep}{0.8em}
\begin{tabular}{cccccc}
\hline
\multicolumn{2}{c}{\underline{     xx     }} & \multicolumn{2}{c}{\underline{     yy     }} & \multicolumn{2}{c}{\underline{     zz     }} \\
      A & 2.2 & I & 2.6 & P & 1.4 \\
      B & 1.6 & J & 2.1 & Q & 7.8 \\
      C & 7.5 & K & 7.4 & R & 2.9 \\
      D & 3.2 & L & 3.3 & S & 4.6 \\
      E & 3.0 & M & 3.3 & T & 4.4 \\
      F & 4.5 & N & 4.4 & U & 3.9 \\
      G & 4.3 & O & 4.1 & - & - \\
      H & 3.5 & - & --  & - & - \\
\hline
\end{tabular}}
\end{center}
\label{VIII}
\end{table}
among all these transitions is the O(p)$\rightarrow$Re(d). All other allowed transitions are presented in table \ref{VIII}. The real value of the dielectric constant at zero energy along different directions are $\epsilon_{1}^{XX}(0)=2.8$, $\epsilon_{1}^{YY}(0)=2.9$, and $\epsilon_{1}^{ZZ}(0)=2.7$ and the corresponding refractive indexes are $n_{XX}(0)=1.7$, $n_{YY}(0)=1.71$, and $n_{ZZ}(0)=1.63$. These values of
\begin{figure}
 \centering
 \includegraphics[width=4in]{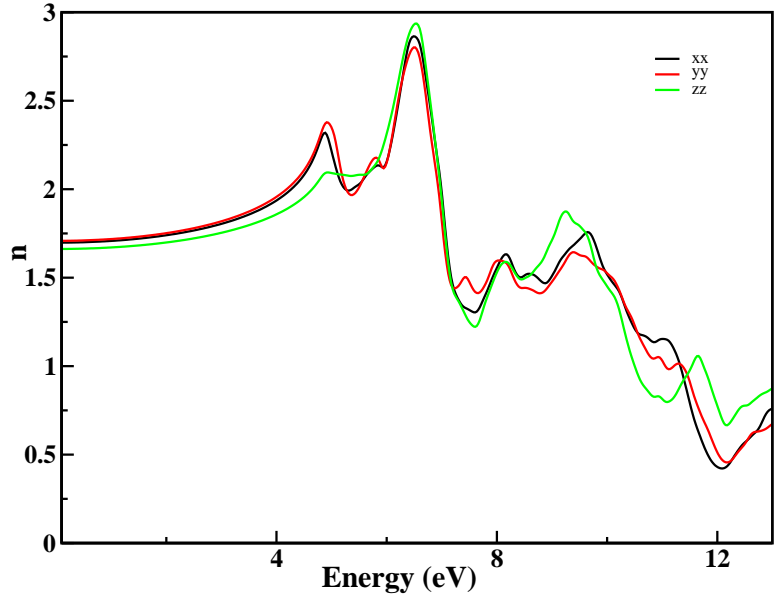}\\
 \caption{(color online) Index of refraction of TlReO$_4$ using experimental crystal structure.}
 \label{8}
\end{figure}
the refractive index undoubtedly show that this parameter along ZZ axis is low whereas is high along the YY axis. Figure \ref{8} shows the three index  of refraction along XX, YY, and ZZ directions. It can be seen from this figure that the maximum value of this index of refraction along XX direction
\begin{figure}
 \centering
 \includegraphics[width=4in]{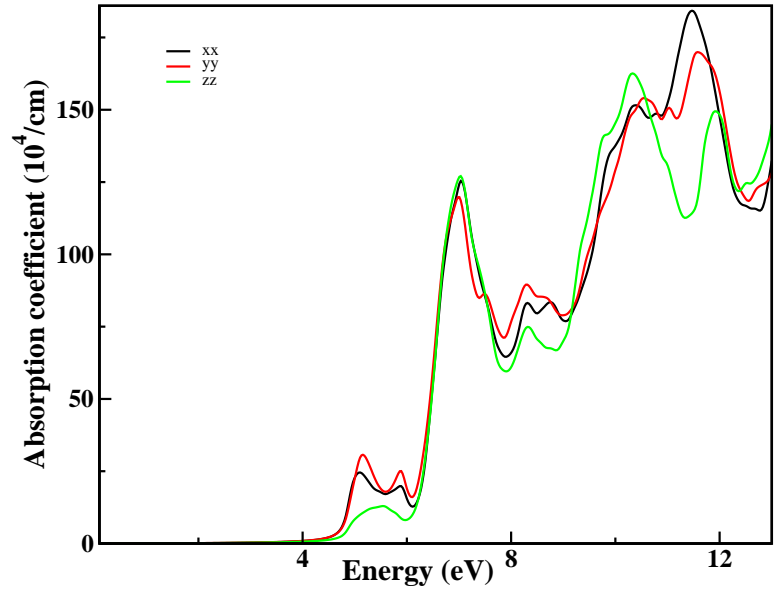}\\
 \caption{(color online) Absorption spectra of TlReO$_4$ using experimental crystal structure.}
 \label{9}
\end{figure}
is 2.9 at 6.5 eV, along YY direction is 2.6 at 6.5 eV and along ZZ direction is 2.9 at 6.5 eV. The absorption spectra of this material is illustrated in figure \ref{9}. It shows the optical isotropy for the energy range 6 to 8 eV and approximately near to 5 eV the first peak has been observed.

\section{Conclusions}
In order to conclude the present work, we have carried out a systematic study of the main physical properties of orthorhombic TlReO$_4$ with the LDA functional. We have obtained and explained the ground state structural and elastic properties of this material. Our calculated bulk modulus is found to be consistent with the reported experimental result and the elastic constant values show the presence of a mechanical anisotropy. This has been interpreted as due to different types of oxygen atoms in different arrangements. After that, the BECs of this material have been studied which clearly provide evidence about the importance of the O3 oxygen atoms present in this material. Zone centered phonon modes have been calculated and assigned. They are presented into three regions to have a better resolution of vibration properties. The strongest peak in the low-frequency range results from the lattice vibration of Tl atom and the ReO$_4$ group in the opposite direction and occurs at 60.74 cm$^{-1}$. The highest intense peak in the 240 to 360 cm$^{-1}$ frequency range is found at 299.15 cm$^{-1}$ and it is due to the scissoring of Re$-$O bonds in ReO$_4$ group. The most intense peak above 900 cm$^{-1}$ is found to occur at 942.07 cm$^{-1}$ being associated to the asymmetric stretching of Re$-$O3 bonds in ReO$_4$ group. Next, the electronic band structure calculation shows that a TB-mBJ potential including a SO coupling properly describes the band gap, which is underestimated by LDA calculations. A study of the density of states established that the main contribution in valence band top comes from O(\emph{p}-states) whereas the conduction band bottom is mainly made up of Re(\emph{d}-states). Though a little amount of isotropy is present in the energy range 6 to 8 eV, the optical properties of TlReO$_4$ clearly suggest that this material is optically anisotropic in nature. In summary, this study clearly explains the structural, mechanical, vibrational, electronic and optical properties of TlReO$_4$. Thorough study of all these properties clearly indicates the importance of O3 oxygen atoms which are aligned in the y-axis and we can conclude that these atoms will carry out an important role in phase transition of this material upon the application of pressure.

\section{Acknowledgments}
S. M. is very much grateful for the financial support provided by DRDO, India, through ACRHEM. Authors also would like to thank CMSD, University of Hyderabad, for helping with computing facilities. D.E. thanks the financial support from the Spanish Ministerio de Ciencia, Innovaci\'{o}n y Universidades, the Spanish Research Agency, the Generalitat Valenciana, and the European Fund for Regional Development under Grants No. MAT2016-75586-C4-1-P and Prometeo/2018/123 (EFIMAT).


\begin{thebibliography}{plain}
\bibitem{1}
G. Voicu, M. Bardoux, and R. Stevensen, Ore Geol. Rev. \textbf{18}, 211 (2001).

\bibitem{2}
M. Nikl, P. Bohacek, N. Mihokova, N. Solovieva, A. Vedda, M. Martini, G. P. Pazzi, P. Fabeni, M. Kobayashi, and M. Ishii, J. Appl. Phys. \textbf{91}, 5041 (2002).

\bibitem{3}
A. Brenier, G. Jia, and C. Tu, J. Phys.: Condens. Matter \textbf{16}, 9103 (2004).

\bibitem{4}
M. Kobayashi, M. Ishii, Y. Usuki, and H. Yahagi, Nucl. Instrum. Methods Phys. Res. A \textbf{333}, 429 (1993).

\bibitem{5}
N. Faure, C. Borel, M. Couchaud, G. Basset, R. Templier, and C. Wyon, Appl. Phys. B \textbf{63}, 593 (1996).

\bibitem{6}
P. R\"{o}gner and K. J. Range, Z. anorg. allg. Chem. \textbf{619}, 1017 (1993).

\bibitem{7}
J. Beintema, Z. Kristallogr.-Cryst. Mater. \textbf{97}, 300 (1937).

\bibitem{8}
L. E. Depero, and L. Sangaletti, J. Solid State Chem. \textbf{129}, 82 (1997).

\bibitem{9}
D. Errandonea, and F. J. Manj\'{o}n, Prog. Mater. Sci. \textbf{53}, 711 (2008).

\bibitem{10}
G. Shwetha, V. Kanchana, K. Ramesh Babu, G. Vaitheeswaran, and M. C. Valsakumar, J. Phys. Chem. C \textbf{118}, 4325 (2014).

\bibitem{11}
D. Errandonea, and A. B. Garg, Prog. Mater. Sci. \textbf{97}, 123 (2018).

\bibitem{12}
G. Shwetha, V. Kanchana, and G. Vaitheeswaran, Materials Chemistry and Physics \textbf{163}, 376 (2015).

\bibitem{13}
S. Mondal, S. Appalakondaiah, and G. Vaitheeswaran, J. Appl. Phys. \textbf{119}, 085702 (2016).

\bibitem{14}
P. K. Jharapla, E. Narsimha Rao, and G. Vaitheeswaran, J. Phys.: Condens. Matter \textbf{30}, 475402 (2018).

\bibitem{15}
O. Fukunaga, and S. Yamaoka, Phys. Chem. Miner. \textbf{5}, 167 (1979).

\bibitem{16}
A. Jayaraman, B. Batlogg, and L. G. VanUitert, Phys. Rev. B \textbf{31}, 5423 (1985).

\bibitem{17}
A. Jayaraman, B. Batlogg, and L. G. VanUitert, Phys. Rev. B \textbf{28}, 4774 (1983).

\bibitem{18}
Von K. Ulbricht, and H. Kriegsmann, Z. Anorg. Allg. Chem. \textbf{358}, 193 (1968).

\bibitem{19}
L. C. Ming, A. Jayaraman, S. R. Shieh, and Y. H. Kim, Phys. Rev. B \textbf{51}, 12100 (1995).

\bibitem{20}
A. Jayaraman, G. A. Kourouklis, R. M. Fleming, and L. G. Van Uitert, Phys. Rev. B \textbf{37}, 664 (1988).

\bibitem{21}
J. M. Ablett, C. C. Kao, S. R. Shieh, H. -K. Mao, M. Croft, and T. A. Tyson, High Pressure Research \textbf{23}, 471 (2003).

\bibitem{22}
A. Jayaraman, G. A. Kourouklis, and L. G. Van Uitert, Phys. Rev. B \textbf{36}, 8547 (1987).

\bibitem{23}
D. Errandonea, A. Mu\={n}oz, P. R. Hern\'{a}ndez, J. E. Proctor, F. Sapi\={n}a, and M. Bettinelli, Inorg. Chem. \textbf{54}, 7524 (2015).

\bibitem{24}
D. de Waal, and W. Kiefer, Z. Anorg. Allg. Chem. \textbf{630}, 127 (2004).

\bibitem{25}
A. Jayaraman, G. A. Kourouklis, L. G. Van Uitert, W. H. Grodkiewicz, and R. G. Maines, Sr., Physica A \textbf{156}, 325 (1988).

\bibitem{26}
P. Hohenberg, and W. Kohn, Phys. Rev. \textbf{136}, B864 (1964).

\bibitem{27}
W. Kohn, and L. J. Sham, Phys. Rev. \textbf{140}, A1133 (1965).

\bibitem{28}
M. C. Payne, M. P. Teter, D. C. Allan, T. A. Arias, and J. D. Joannopoulos, Rev. Mod. Phys. \textbf{64}, 1045 (1992).

\bibitem{29}
M. D. Segall, P. J. D. Lindan, M. J. Probert, C. J. Pickard, P. J. Hasnip, S. J. Clark, and M. C. Payne, J. Phys.: Condens. Matter. \textbf{14}, 2717 (2002).

\bibitem{30}
D. M. Ceperley, and B. J. Alder, Phys. Rev. Lett. \textbf{45}, 566 (1980).

\bibitem{31}
J. P. Perdew, and A. Zunger, Phys. Rev. B \textbf{23}, 5048 (1981).

\bibitem{32}
H. J. Monkhorst, and J. D. Pack, Phys. Rev. B \textbf{13}, 5188 (1976).

\bibitem{33}
P. Blaha, K. Schwarz, G. K. H. Madsen, D. Kvasnicka, J. Luitz, ISBN 3-9501031-1-2, (2001).

\bibitem{34}
F. Tran, and P. Blaha, Phys. Rev. Lett. \textbf{102}, 226401 (2009).

\bibitem{35}
F. D. Murnaghan, Proc. Natl. Acad. Sci. U. S. A. \textbf{30}, 244 (1944).

\bibitem{36}
M. Posternak, R. Resta, and A. Baldereschi, Phys. Rev. B \textbf{50}, 8911 (1994).

\bibitem{37}
W. Zhong, R. D. King-Smith, and D. Vanderbilt, Phys. Rev. Lett. \textbf{72}, 3618 (1994).

\bibitem{38}
D. Errandonea, J. P. Porres, F. J. Manj\'{o}n, A. Segura, Ch. F. Roca, R. S. Kumar, O. Tschauner, P. R. Hern\'{a}ndez, J. L. Solano, S. Radescu, A. Mujica, A. Mu\~{n}oz, and G. Aquilanti, Phys. Rev. B \textbf{72}, 174106 (2005).

\bibitem{39}
V. Panchal, N. Garg, H. K. Poswal, D. Errandonea, P. R. Hern\'{a}ndez, A. Mu\~{n}oz, and E. Cavalli, Phys. Rev. Materials \textbf{1}, 043605 (2017).

\bibitem{40}
D. Errandonea, R. S. Kumar, X. Ma, C. Tu, J. Solid state Chem \textbf{181}, 355 (2008).

\bibitem{41}
H. Kunkely, and A. Vogler, Monatshefte f\"{u}r Chemie \textbf{135}, 1 (2004).
\end{thebibliography}
\end{document}